\pdfoutput=1
\documentclass[a4paper,fleqn,usenatbib]{mnras}

\usepackage[T1]{fontenc}
\usepackage{ae,aecompl}

\usepackage{graphicx}	
\usepackage{savesym}
\usepackage{amsmath}
\savesymbol{iint}
\usepackage{txfonts}
\restoresymbol{TXF}{iint}
\usepackage{amssymb}	
\usepackage[usenames,dvipsnames]{color}
\usepackage[normalem]{ulem}
\usepackage{lipsum}
\usepackage{multirow,array}
\usepackage{hhline}
\usepackage[para,online,flushleft]{threeparttable}
\usepackage{hyperref}

\title[Faint Dwarf Galaxies in HCG90]{Faint Dwarf Galaxies in Hickson Compact Group 90\thanks{Based on observations collected under program 094.B-0366 at the Very Large Telescope of the Paranal Observatory in Chile, operated by the European Southern Observatory (ESO).}}

\author[Ordenes-Brice\~no et al.]{Yasna Ordenes-Brice\~no$^{1,}$\thanks{E-mail:yordenes@astro.puc.cl}, Matthew A.~Taylor$^{1,2}$, Thomas H.~Puzia$^{1}$, \newauthor
Roberto P.~Mu\~noz$^{1}$, Paul Eigenthaler$^{1}$, Iskren Y. Georgiev$^{3}$, Paul Goudfrooij$^{4}$, \newauthor
Michael Hilker$^{5}$, Ariane Lan\c{c}on$^{6}$, Gary Mamon$^{7}$, Steffen Mieske$^{8}$, \newauthor
Bryan W. Miller$^{9}$, Eric W. Peng$^{10,11}$, \& Rub\'en S\'anchez-Janssen$^{12}$
\\
$^{1}$Institute of Astrophysics, Pontificia Universidad Cat\'olica de Chile, Av.~Vicu\~na Mackenna 4860, 7820436 Macul, Santiago, Chile\\
$^{2}$European Southern Observatory, Alonso de Cordova 3107, Vitacura, Santiago, Chile\\
$^{3}$Max-Planck Instiut f\"ur Astronomie, K\"onigstuhl 17, D-69117 Heidelberg, Germany\\
$^{4}$Space Telescope Science Institute, 3700 San Martin Drive, Baltimore, MD 21218, USA\\
$^{5}$European Southern Observatory (ESO), Karl-Schwarzschild-Strasse 2, D-85748 Garching, Germany\\
$^{6}$Observatoire Astronomique de Strasbourg, Universit\'e de Strasbourg, CNRS, UMR 7550, 11 rue de l'Universit\'e , 67000, Strasbourg, France\\
$^{7}$Institut d'Astrophysique de Paris (UMR 7095: CNRS and UPMC, Sorbonne Universit\'es), F-75014 Paris, France\\
$^{8}$European Southern Observatory, 3107 Alonso de C\'ordova, Vitacura, Santiago\\
$^{9}$Gemini Observatory, Casilla 603, La Serena, Chile\\
$^{10}$Department of Astronomy, Peking University, Beijing 100871, China\\
$^{11}$Kavli Institute for Astronomy \& Astrophysics, Peking University, Beijing 100871, China\\
$^{12}$NRC Herzberg Astronomy and Astrophysics, 5071 West Saanich Road, Victoria, BC V9E 2E7, Canada
}

\date{Accepted 12 August, 2016. }

\pubyear{2016}

\begin{document}
\label{firstpage}
\pagerange{\pageref{firstpage}--\pageref{lastpage}}
\maketitle

\begin{abstract}
We report the discovery of a very diverse set of five low-surface brightness (LSB) dwarf galaxy candidates in Hickson Compact Group 90 (HCG\,90) detected in deep $U$- and $I$-band images obtained with VLT/VIMOS. These are the first LSB dwarf galaxy candidates found in a compact group of galaxies.~We measure spheroid half-light radii in the range $0.7\!\lesssim\! r_{\rm eff}/{\rm kpc}\! \lesssim\! 1.5$ with luminosities of $-11.65\!\lesssim\! M_U\! \lesssim\! -9.42$ and $-12.79\!\lesssim\! M_I\! \lesssim\! -10.58$\,mag, corresponding to a color range of $(U\!-\!I)_0\!\simeq\!1.1\!-\!2.2$\,mag and surface brightness levels of $\mu_U\!\simeq\!28.1\,{\rm mag/arcsec^2}$ and $\mu_I\!\simeq\!27.4\,{\rm mag/arcsec^2}$.~Their colours and luminosities are consistent with a diverse set of stellar population properties.~Assuming solar and $0.02\,Z_\odot$ metallicities we obtain stellar masses in the range $M_*|_{Z_\odot}\!\simeq\!10^{5.7-6.3}\,M_{\odot}$ and  $M_*|_{0.02\,Z_\odot}\!\simeq\!10^{6.3-8}\,M_{\odot}$.~Three dwarfs are older than 1\,Gyr, while the other two significantly bluer dwarfs are younger than $\sim\!2$\,Gyr at any mass/metallicity combination.~Altogether, the new LSB dwarf galaxy candidates share properties with dwarf galaxies found throughout the Local Volume and in nearby galaxy clusters such as Fornax.~We find a pair of candidates with $\sim\!2$\,kpc projected separation, which may represent one of the closest dwarf galaxy pairs found.~We also find a nucleated dwarf candidate, with a nucleus size of $r_{\rm eff}\!\simeq\!46\!-\!63$\,pc and magnitude M$_{U,0}=-7.42$\,mag and $(U\!-\!I)_0\!=\!1.51$\,mag, which is consistent with a nuclear stellar disc with a stellar mass in the range $10^{4.9-6.5}\,M_\odot$.
\end{abstract}




\section{Introduction}
In recent years, deep observations conducted with wide-field imaging cameras are reaching low-surface brightness (LSB) levels, comparable with the faint outer stellar halos of galaxies ($\mu_i\!\approx\!28-30$\,mag arcsec$^{-2}$).~Many LSB dwarf and/or ultra-diffuse galaxies (UDGs) with $\langle\mu_i\rangle_{\rm eff}\!\geq\!25$\,mag arcsec$^{-2}$ have been identified in a range of environments including the Local Group \cite[LG; see e.g.][]{McCon12,koposov15}, nearby galaxies \citep[e.g.][]{java16}, loose galaxy groups like Centaurus A \citep[e.g.][]{crnojevic15}, and in more massive and dense galaxy clusters like Fornax, Virgo, and Coma \citep[e.g.][]{munoz15,mihos15,vandok15,koda15}.~Such dense environments where galaxy-galaxy interactions are relatively common are interesting sites to study dwarfs due to their susceptibility to galaxy transformation processes \citep[e.g.][]{Lisker09,janz12,rys14,rys15}.

The varied environments where faint dwarf galaxies have been detected so far make compact galaxy groups tempting places to look for LSB dwarfs and UDGs.~Hickson compact groups (HCGs) are defined as small groups of four or more massive galaxies located in relative isolation \citep{hick97}.~Given that they rival the cores of galaxy clusters as the most dense galaxy environments in the nearby universe \cite[e.g.][]{rubin91,ponman96,proctor04}, and with typical velocity dispersion of $\sigma\!\simeq\!200\,{\rm km}\,{\rm s}^{-1}$ \citep{hick92}, HCGs are expected to be the ideal environment for galaxy mergers and tidal interactions \citep[e.g.][]{mamon92}.~Such events can also generate kinematically decoupled structures with active star formation, so-called tidal dwarf galaxies (TDGs), which have masses similar to dwarf galaxies resembling the metallicities of their hosts and lacking dark matter  \citep[e.g.][]{kroupa10,gall10}. HCGs may also host UDGs and constitute analogues to sites of dwarf galaxy pre-processing before their infall onto larger clusters.

At a distance of 33.1\,Mpc \citep[$m\!-\!M\!=\!32.6$\,mag, see][]{bla01}, HCG\,90 is one of the most nearby compact galaxy groups accessible from the Southern Hemisphere.~The 19 known members have a group radial velocity of $v_r\!\simeq\!2\,600\,{\rm km}\,{\rm s}^{-1}$, and a velocity dispersion of $\sigma\!\simeq\!193\,{\rm km}\,{\rm s}^{-1}$, typical of HCGs \citep{zablu98}.~The core is dominated by three bright galaxies (NGC\,7173/HCG\,90b, NGC\,7176/HCG\,90c, NGC\,7174/HCG\,90d), with a fourth giant galaxy (NGC\,7172/HCG\,90a) located to the north of the core (see Fig.\,\ref{fig:members}), which is a Seyfert\,2 galaxy and a bright X-ray source.~Despite the relative proximity, there have been no detailed studies on its LSB dwarf galaxy population, as most attention has been paid to the group's bright and/or ultra-compact dwarf (UCD) \citep{car94,rib94,dar11}, and giant galaxies.~This avenue of study is ripe for investigation, as other HCGs have been shown to host faint dwarfs \citep{cam04,car06,kru06,kon13}, which are to be expected from theoretical considerations, despite the dynamically more hostile environments in these systems \citep{zan14}.

HCG\,90 is at an interesting stage of evolution.~Several indications of interactions between its galaxies are evidenced by morphological disturbances, a diffuse X-ray halo, and intra-group light \citep{oliv94,zablu98,White03,Desja13}.~It is proposed that current interactions between the core galaxies, strongly between HCG\,90b/d and weakly between HCG\,90c/d, have given rise to diffuse intra-group light that contributes $\sim\!45$ percent to the total light \citep{White03}.~Additionally, a warm gas envelope is shared by HCG\,90b and d, while tidal bridges have been found between HCG\,90b, c, and d \citep{plana98,oliv94}.~All this makes HCG\,90 an intriguing laboratory to investigate interactions between giant and dwarf galaxies, and their star cluster satellites.~In this work we report the discovery of five LSB dwarf galaxy candidates associated with HCG\,90 based on deep $U$- and $I$-band imaging, and compare their properties with those of other LSB dwarf galaxies recently reported in the literature.

\section{Observations and Data Reduction}\label{Sect:Data}
\begin{figure}
\includegraphics[width=8.2cm,height=8cm]{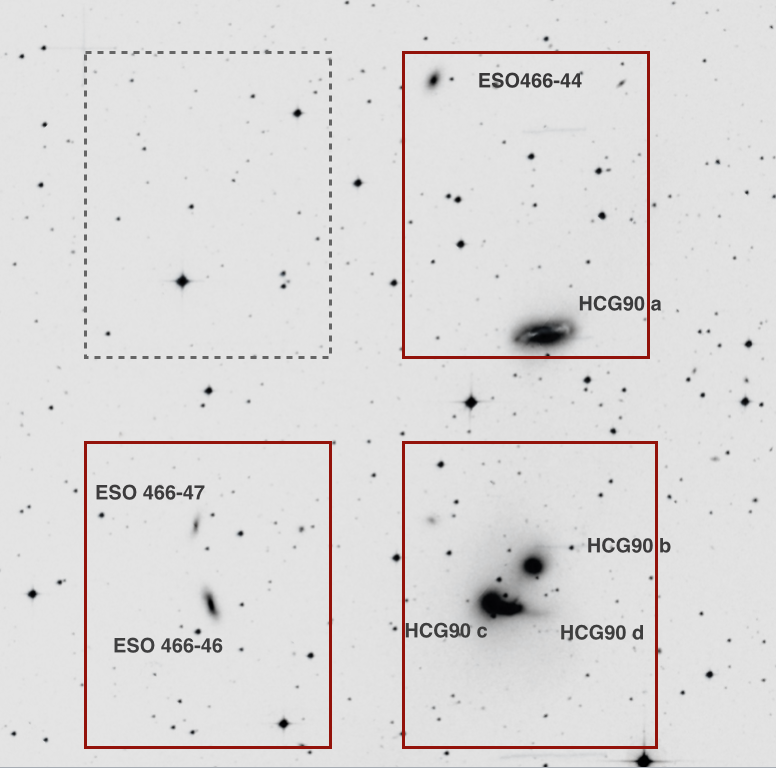}
\caption{SDSS colored image of the central $21\arcmin\!\times\!21\arcmin$ region in HCG\,90, corresponding to 202\,kpc\,$\times\,$202\,kpc assuming a distance of 33.1\,Mpc.~Several prominent group members are  indicated on the figure.~The rectangles show the field of view of VLT/VIMOS.~Note that due to severe guide-probe vignetting during the service-mode observations we were not able to use the upper left quadrant for our analysis (dashed rectangle).\label{fig:members}}
\end{figure}

We obtained deep near-ultraviolet (NUV) and optical imaging with the {\it Visible Multi-Object Spectrograph} (VIMOS) instrument in the $U$- and $I$-bands as part of the service-mode programme P94.B-0366 (PI: Taylor).~VIMOS is mounted on UT3 (Melipal) of the {\it Very Large Telescope} (VLT), and is comprised of four CCDs each with a $7\arcmin\times8\arcmin$  field of view and a pixel size of 0.205"$\simeq$\,37.3\,pc at a distance of 33.1 Mpc.~The four VIMOS quadrants were placed to cover the four giant galaxies HCG\,90a, b, c, and d, as well as several of the known group galaxies (see Fig.\,\ref{fig:members}).

The NUV observations were broken up into five observing blocks (OBs) each consisting of $4\times930\,{\rm s}$ exposures, conducted during the nights of October 24-25th, 2014.~An additional OB with $3\times930\,{\rm s}$ was observed on November 18th, 2014; however, we do not include these data in the present work due to the significantly poorer seeing ($\sim\!1.8\arcsec$) compared to the previous observations ($0.7\!-\!1.2\arcsec$).~The optical images were likewise taken under excellent seeing conditions ($0.6\!-\!1.0\arcsec$), and were broken up into two OBs each of $6\times546\,{\rm s}$ exposures conducted on October 1st and 28th, 2014.~All sub-integrations were dithered with steps of a few arcseconds.~One quadrant of the VIMOS field of view was significantly vignetted by the guide probe and thus not used in the subsequent analysis (see dashed box in Fig.\,\ref{fig:members}).

\begin{figure*}
    \includegraphics[trim=3cm 2cm 7cm 1cm,clip=true,scale=0.3305]{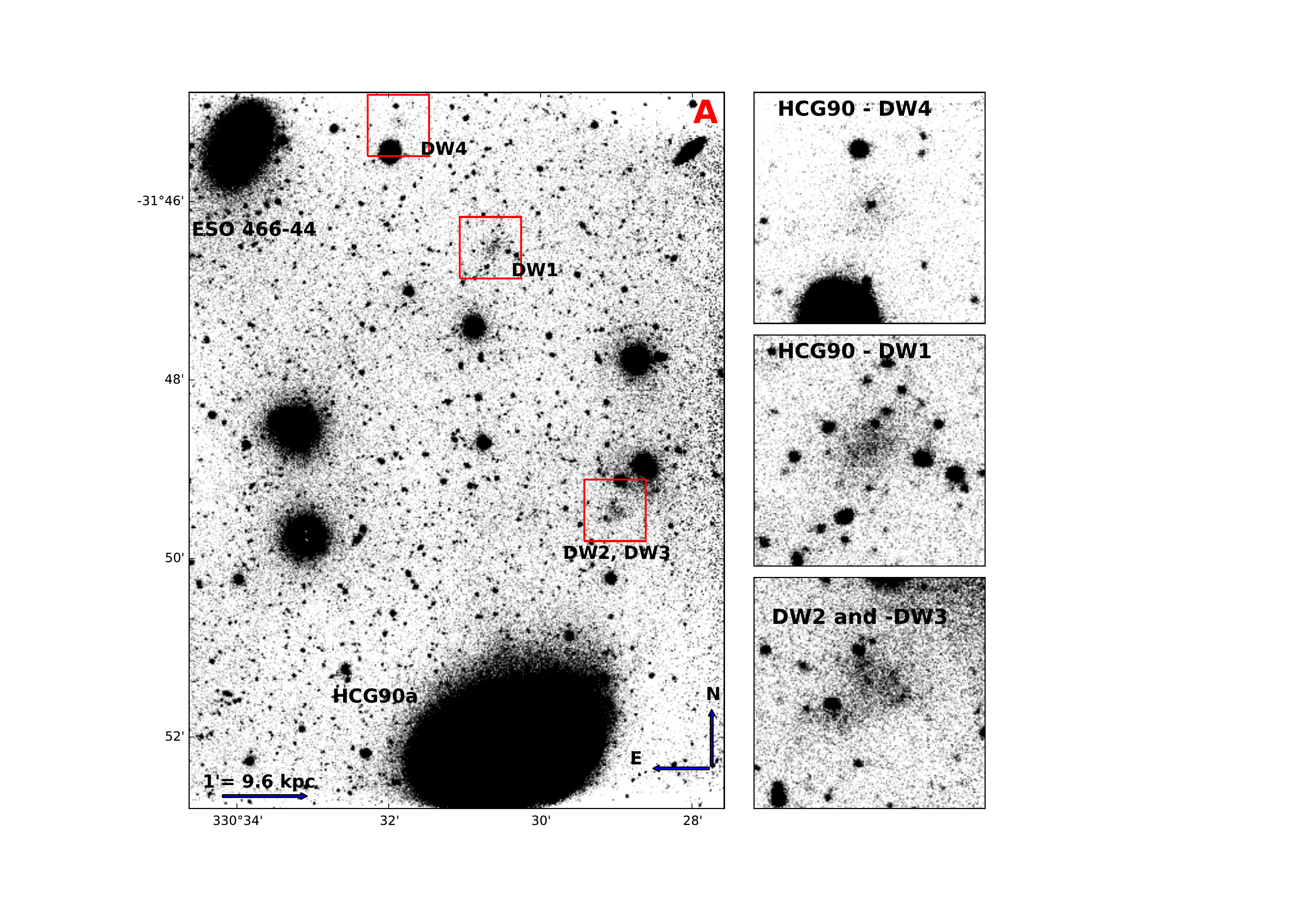}\includegraphics[trim=3.5cm 2cm 7cm 1cm,clip=true,scale=0.37]{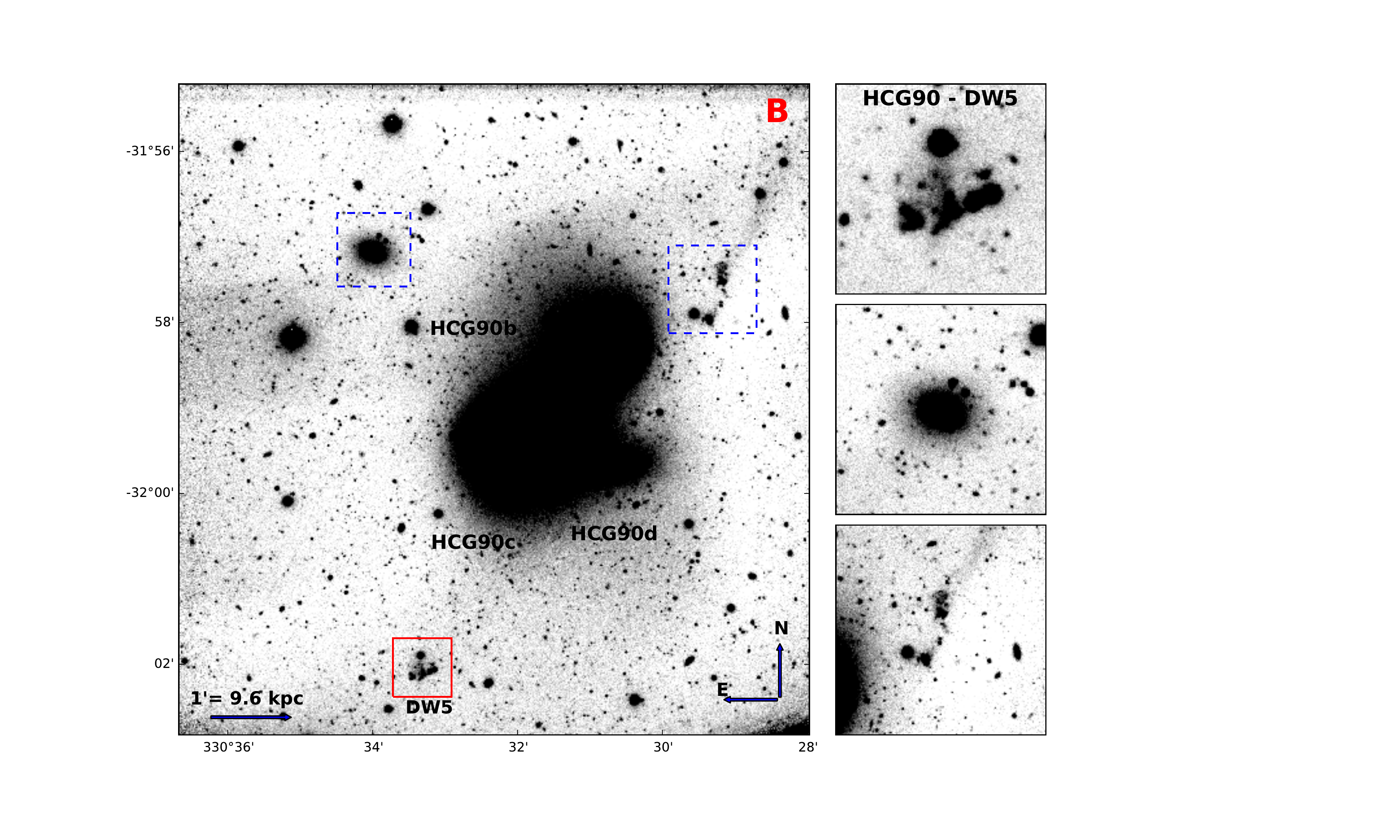}
     \caption{$U$-band VLT/VIMOS images of the new dwarf candidates, marked by red boxes.~{\it Panel A:}\,The region around HCG90a and the galaxy ESO466-44 is shown alongside candidate postage-stamp cutouts of the individual dwarf candidates.~{\it Panel B:}\,A similar view of the core region of HCG\,90 is shown.~A new dwarf galaxy candidate is indicated by the red box, while blue dashed boxes indicate a member previously catalogued by \protect\cite{zablu98}, as well as a candidate potentially caught in the process of being tidally disrupted by HCG\,90b.\label{fig:dwarfs}}
\end{figure*}
\begin{figure*}
   \includegraphics[height=0.195\linewidth]{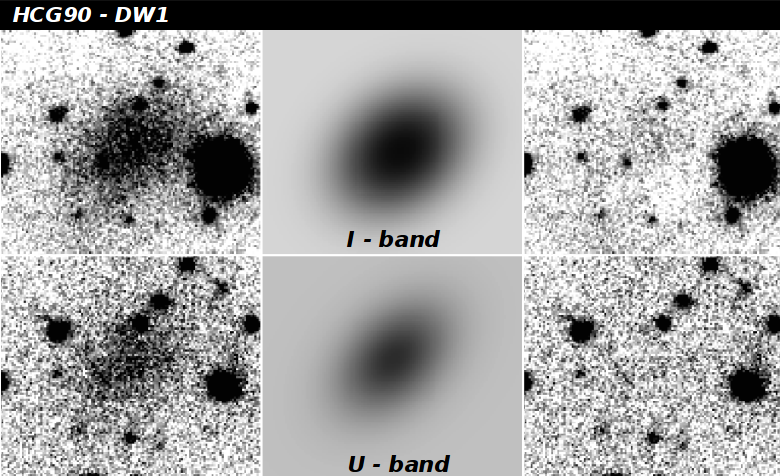}
   \includegraphics[height=0.195\linewidth]{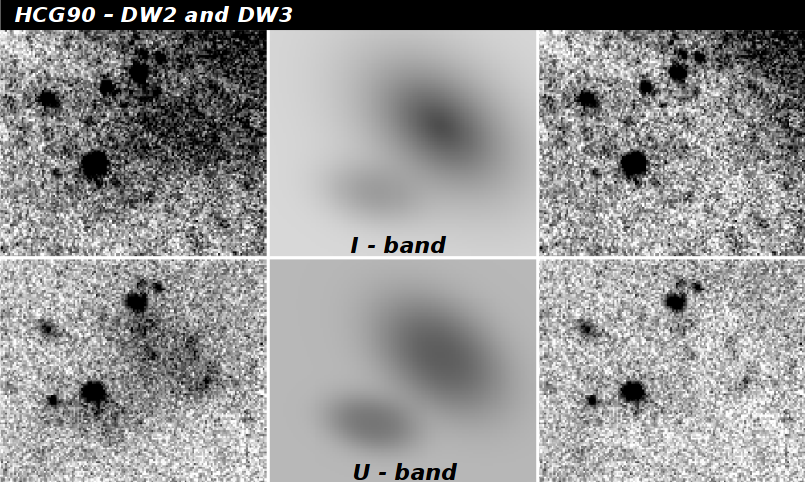}
   \includegraphics[height=0.195\linewidth]{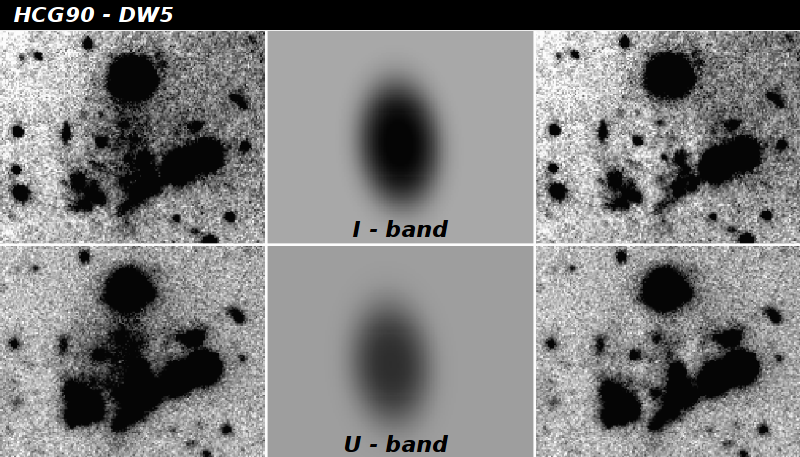}
   \caption{Illustrations of the surface brightness light profile fitting procedure, conducted in both $I$- and $U$-bands, shown in the top and bottom panel rows, respectively.~For each of the three panels, the left-most postage stamps are the processed science frames in the corresponding filter, while the middle and right-most cutouts show our {\sc galfit} models and residual images, respectively.\label{fig:dwarf_model}}
\end{figure*}

The raw images were processed by the VIMOS pipeline (v3.0.6) using the {\sc esorex} (v3.12) framework to correct for bias, flat-fielding, bad pixels, and cosmic rays.~The background (sky) subtraction and final image stacking were performed using custom {\sc Python} scripts.~For the background subtraction, we mask objects in the field, and model the sky with a thin plate spline rather than using contiguous images to estimate the sky level.~To register and stack the individual exposures we use the {\sc astromatic}\footnote{http://www.astromatic.net} software {\sc SExtractor} \citep[v2.19.5;][]{bertin96}, {\sc scamp} \citep[v2.0.4;][]{bertin06}, and {\sc SWarp} \citep[v2.38.0;][]{bertin02}.~For the astrometric calibration we use reference stars from the 2MASS Point Source Catalog \citep{Skrutskie06}.~Photometric zero points are calculated based on standard star fields from the Stetson catalog \citep{stet00} taken during the observations, which were processed in an identical way as the science frames.

\section{Image Analysis}\label{Sect:analysis}
The final $U$- and $I$-band stacks were visually inspected for diffuse sources characteristic of dwarf galaxies, revealing five potential candidates.~The red boxes in Fig.\,\ref{fig:dwarfs} show the locations of the new dwarfs, while blue boxes show a dwarf galaxy previously discovered by \cite{zablu98} and tidal debris associated with HCG\,90b.~Four of the new dwarf candidates, HCG90-DW1, -DW2, -DW3 and -DW4 (see Fig.\,\ref{fig:dwarfs}, panel {\it A}), are located between HCG\,90a and ESO\,466-44, while the fifth dwarf candidate, HCG90-DW5, is located to the SE of the three central HCG\,90 galaxies (see panel {\it B}).~Small postage-stamp panels show more detailed views of the new dwarf galaxies.~From top-to-bottom in panel\,{\it A} we note a potentially nucleated dwarf candidate with a compact central component, a LSB candidate with a relatively undisturbed morphology, and lastly, a candidate with a very irregular shape.~The latter object shows two components that potentially represent a binary dwarf system similar to those identified in more diffuse galaxy groups \citep[e.g.][]{crno14}.

The top postage stamp beside panel\,{\it B} shows the fifth dwarf candidate, which is unfortunately heavily contaminated by foreground stars that prevent accurate determination of its structural parameters.~To the East of HCG\,90b, we locate an object catalogued by \cite{zablu98} as a member of HCG\,90 which is likely to be a dwarf galaxy; however, there are no further studies of this object in the literature.~In the bottom postage-stamp panel, we show tidal features associated with the galaxy HCG\,90b.~This feature shows signs of an object caught in the act of disruption and/or tidal stripping.~For the five new dwarf candidates, we derive structural parameters and investigate their morphological properties.

\begin{figure*}
\includegraphics[width=\linewidth]{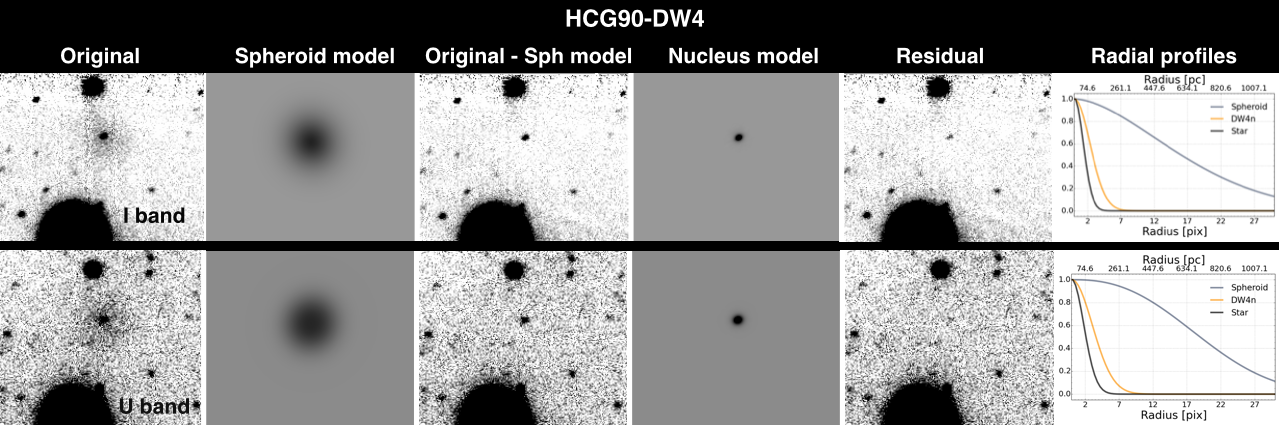}
\caption{Illustrations of the surface brightness profile fitting for the spheroid and central object of HCG90-DW4 in $I$ and $U$-band (top and bottom panel rows, respectively).~The original image is shown in the left-most panel, then to the right follows the spheroid model with a spheroid-subtracted image. The following two panels show the nucleus model and the residual, where spheroid+nucleus model was subtracted. In the right-most panel we show the radial profiles of the spheroid S\'ersic profile, the nucleus DW4n, and a stellar point spread function.~Note that the nucleus, DW4n, is resolved and slightly more extended that the stellar profile.~The image panels for each filter are shown with a greyscale stretch that is equal for each panel.\label{fig:dwarf4}}
\end{figure*}

\begin{table*}
\caption{HCG\,90 LSB dwarf candidate photometry and structural parameters.~The last row summarizes the corresponding information for the central object in HCG90-DW4 performed with S\'ersic surface brightness model.\label{table:info}}
\begin{threeparttable}
\centering
\setlength{\tabcolsep}{3pt} 
\renewcommand{\arraystretch}{1.2} 
\begin{tabular}{cccccccccccc}
\hline
\hline
ID         & RA(J2000)     & Dec(J2000)      & $M_{U,{0}}$ & $n_{\,U}$ $^{a}$ &$r_{\rm eff_{U}}$ &$r_{\rm eff_{\,U}}$ $^{b}$ & $b/a$ & $n_{\,I}$ $^{a}$ &$r_{\rm eff_{\,I}}$ &$r_{\rm eff_{\,I}}$ $^{b}$ & $(U-I)_{0}$ \\
           & [hh:mm:ss.ss] & [dd:mm:ss.ss] & [mag]  &     & [arcsec] &  [kpc] &   &     & [arcsec]& [kpc] & [mag] \\
\hline
\vspace{0.13cm}
HCG90-DW1 &  22:02:02.47  &  $-$31:46:31.87  & $-$10.50 $^{+0.02} _{-0.09}$& 0.51 & 5.27 $^{+0.06} _{-0.08}$ & 0.959 $^{+0.011} _{-0.014}$& 0.67 & 0.42 & 5.53 $^{+0.12} _{-0.01}$& 1.006 $^{+0.021} _{-0.001}$& 2.21\\
\vspace{0.13cm}
HCG90-DW2 &  22:01:55.97  &  $-$31:49:29.46  & $-$10.83 $^{+0.08} _{-0.10}$ & 0.42 & 6.06 $^{+0.40} _{-0.19}$& 1.103 $^{+0.072} _{-0.034}$& 0.65 & 0.57 &   8.34 $^{+0.60} _{-0.55}$& 1.520 $^{+0.108} _{-0.099}$&  1.96  \\
\vspace{0.13cm}
HCG90-DW3 &  22:01:56.58  &  $-$31:49:35.86  & $-$9.42 $^{+0.04} _{-0.03}$  & 0.35 & 3.76 $^{+0.23} _{-0.32}$& 0.684 $^{+0.042} _{-0.058}$& 0.55 & 0.48  &   4.67 $^{+0.74} _{-0.32}$&  0.850 $^{+0.134} _{-0.058}$&  1.16  \\
\vspace{0.13cm}
HCG90-DW4 &  22:02:07.44  &  $-$31:45:07.32  &  $-$10.02 $^{+0.1} _{-0.2}$& 0.41   & 4.11 $^{+0.11} _{-0.22}$ &  0.747 $^{+0.021} _{-0.040}$& 0.95&  0.58  & 4.43 $^{+0.42} _{-0.35}$  & 0.805 $^{+0.076} _{-0.063}$& 2.12   \\
\vspace{0.13cm}
HCG90-DW5 &   22:02:13.41 &  $-$32:02:04.39 & $-$11.65 $^{+0.2} _{-0.2}$& 0.35 & 5.68 $^{+0.31} _{-0.30}$& 1.034 $^{+0.056} _{-0.054}$ & 0.65 &  0.31   &  5.81 $^{+0.33} _{-0.30}$  &   1.057 $^{+0.055} _{-0.060}$  & 1.14 \\
\hline
HCG90-DW4n $^{c}$ &  22:02:07.45  &  $-$31:45:07.12  &  $-$7.42 $^{+0.1} _{-0.2}$  & 0.69   & 0.35 $^{+0.06} _{-0.12}$   &  0.063 $^{+0.010} _{-0.022}$& 0.48 &  0.71   &  0.25 $^{+0.11} _{-0.07}$  &  0.046 $^{+0.020} _{-0.002}$& 1.51\\
\end{tabular}
\begin{tablenotes}
$^{a}$ S\'ersic index \citep{sersic68,caon93}; \\
$^{b}$ Assuming a distance of 33.1\,Mpc \protect\citep{bla01};\\
$^{c}$ HCG90-DW4 central object - S\'ersic profile
\end{tablenotes}
\end{threeparttable}
\end{table*}

The structural parameters for each dwarf candidate are determined using {\sc galfit} \cite[v.3.0.5.;][]{peng10} following a multi-step procedure.~We first create individual $41\arcsec \times 41\arcsec$ ($7.4$\,$\times$\,$7.4$\,kpc) cutout images centered on each candidate.~Next we use the {\sc SExtractor} segmentation maps to mask the neighboring sources to model only the dwarf light.~In addition, an input PSF image is also used, which is created using {\sc PSFex} \cite[v.3.16.1;][]{bertin11}.~The surface brightness distributions are fit with a one-component S\'ersic profile \citep{sersic68}.~As these dwarf candidates are LSB galaxies, {\sc galfit} initially could not find stable fit solutions, except for HCG90-DW1, for which the solution easily converged.~For the rest, we apply the same iterative technique developed for the recently discovered faint dwarf candidates in the Fornax cluster \cite[][see their \S\,3]{munoz15} to refine the fit.~First, we estimate the total galaxy magnitude from {\sc SExtractor} MAG\_AUTO as an initial guess and keep it fixed while fitting the S\'ersic model.~The output parameters from this run are then fixed and the magnitude is recomputed.~Finally, for the last run, we again estimate the parameters keeping the newly measured magnitude fixed.~The resulting models and residuals are visually inspected for each dwarf candidate to confirm that the fits are robust, and we summarize the derived structural parameters in Table\,\ref{table:info}.

\begin{figure*}
    \includegraphics[trim=0.3cm 0.3cm 0.3cm 0.3cm,clip=true,width=8.85cm]{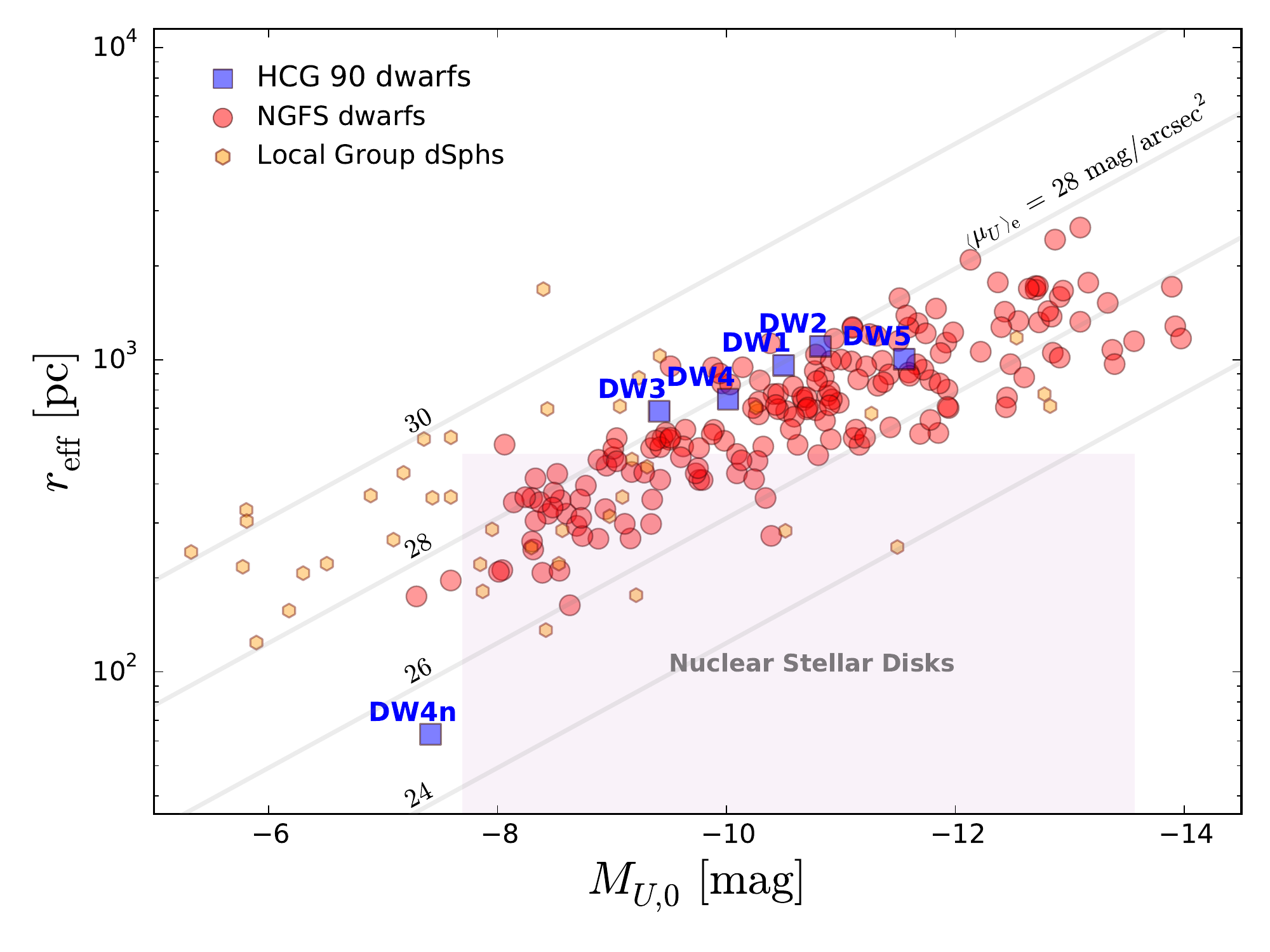}
    \includegraphics[trim=0.1cm 0.1cm 0.3cm 0.0cm,clip=true,width=8.5cm]{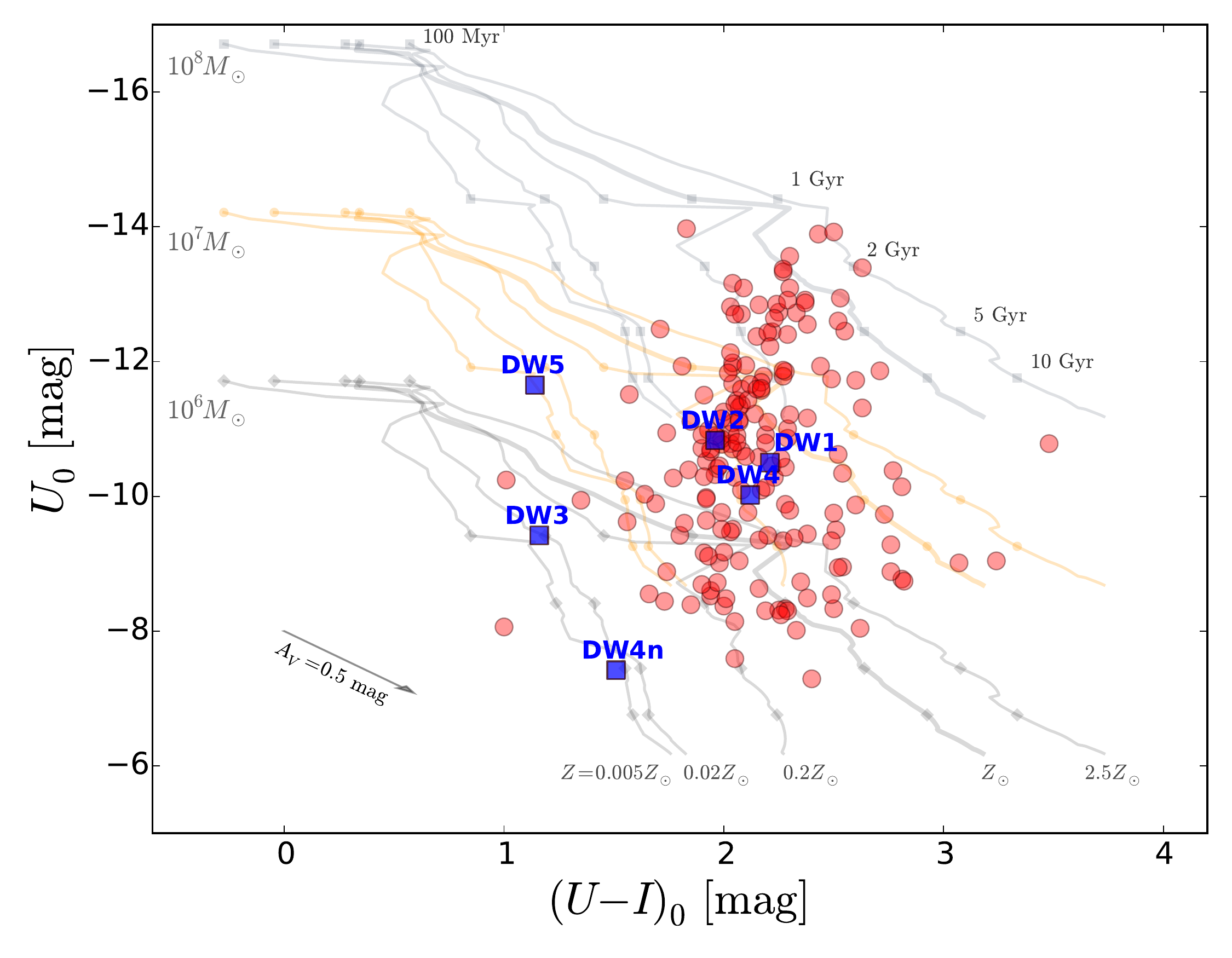}
    \caption{{\it Left panel:}\,The size-luminosity relation for dwarf galaxies.~Effective radii vs. absolute $U$-band magnitude are shown for the LSB dwarf galaxy candidates in HCG\,90 (blue squares).~Lines of constant average effective surface brightness are shown for $\langle\mu _U\rangle_e$=30,\,28,\,26,\,24 mag arcsec$^{-2}$.~Red circles are the Fornax cluster dwarf candidates identified by \protect\cite{munoz15} and hexagons are the LG dSphs \protect\citep{McCon12}.~The shaded area shows the parameter space of known nuclear stellar discs \protect\citep{ledo10}.
{\it Right panel:}\,Color-magnitude diagram, $U_0$ vs $(U\!-\!I)_0$ for the HCG\,90 dwarf candidates (blue squares) and Fornax cluster dwarfs (red circles).~SSP model predictions from \protect\cite{bc03} are shown for three different total stellar masses ($10^6$, $10^7$ and $10^8 M_\odot$) from top to bottom.~Iso-metallicity tracks are shown for Z\,=\,0.05, 0.02, 0.2, 1.0 (thicker line), and 2.5\,Z$_\odot$ and ages older than 100\,Myr.~The arrow in the upper left corner indicates a reddening vector corresponding to $A_V\!=\!0.5$\,mag.\label{fig:lumsize}}
\end{figure*}

Fig.\,\ref{fig:dwarf_model} shows the best fit models derived for four candidates in the $U$- and $I$-bands.~For HCG90-DW4, the spheroid and the nuclear star cluster are treated separately (see Fig.\,\ref{fig:dwarf4} and \S\,\ref{Sect:discussion}).~We point out that the VIMOS PSF is fully accounted for in the surface brightness profile fit of every component, in particular the central object of HCG90-DW4.~We correct the magnitudes for Galactic foreground extinction ($E_{B-V}=0.024$\,mag) using the latest \cite{sch11} recalibration of the \cite{schlegel98} dust reddening maps.~For HCG\,90 the corresponding reddening values are $A_U\!=$\,0.115\,mag and $A_I\!=$\,0.04\,mag estimated with the \cite{fitz99} reddening law ($R_V\!=\!3.1$). In general, the candidates are well characterized in the $U$-band with average effective surface brightness levels of $\langle\mu _U\rangle_e\!\simeq\!26.7\!-\!28.1$\,mag arcsec$^{-2}$, and spheroid effective radii of $r_{\rm eff}\!=\!4\!-\!6\arcsec$ corresponding to $0.68\!-\!1.10$\,kpc.

For the dwarf pair HCG90-DW2 and HCG90-DW3, a single-component S\'ersic profile did not result in a stable solution.~However, we find that a two-component profile with different centers rapidly converges to a stable solution (see center panel in Fig.\,\ref{fig:dwarf_model}).~We thus consider the two components separately.~While both components of HCG90-DW2 and -DW3 are detected and fit in both filters, the saturation of a nearby star prohibits a robust estimation of their structural parameters in the $I$-band.~Fortunately this is not the case for the $U$-band, so that the profile parameters from this filter can be used as first guess for the $I$-band surface brightness distribution fits.~Despite finding a stable fitting solution, we point out that this requires considerable manual interaction with {\sc galfit}.~Our results imply that we are looking either at a dwarf pair (potentially in projection) or an extended dwarf irregular system.~Given that we do not detect increased stochasticity in our $U$-band image that would indicate star-forming regions, with the current information we consider the dwarf pair scenario most likely.

Of the five new dwarf candidates, HCG90-DW4 represents the sole example of a nucleated dwarf galaxy (dE,N; see Fig.\,\ref{fig:dwarf4}).~Its spheroid component has $r_{\rm eff}\!\approx\!750\!-\!800$\,pc (see Table\,\ref{table:info}).~Fitting the central object (DW4n) with a S\'ersic profile yields an $n\!\simeq\!0.6\!-\!0.7$ index, with $r_{\rm eff}\!\simeq\!46\!-\!63$\,pc and a magnitude of $M_{U,{0}}\!=\!-7.42$\,mag.~The corresponding nucleus to spheroid luminosity ratio is ${\cal L}_n/{\cal L}_g\!\simeq\!0.091$, which is significantly above the 0.41 percent found by \cite{tur12} for bright Fornax early-type galaxies, but is in line with the results from the {\it Next Generation Fornax Survey} which found similarly elevated ${\cal L}_n/{\cal L}_g$ ratios for faint dwarf galaxies \citep{munoz15}.~We also find a moderate ellipticity of the central source, but as we are near the resolution limit of our data we cannot draw strong conclusions from this estimate.~A faint residual is left near the centre of the nucleus, visible in the most right-panel of Fig.\,\ref{fig:dwarf4} for both bands.~It is brighter in $I$ than in the $U$-band, with a similar position angle, which might be a potential background source or a dust lane associated with the nucleus itself.~Higher spatial resolution data is required to obtain further information about the morphology of this object.~If DW4n is a nuclear star cluster (NSC), it appears to be larger and fainter than usual, falling well off the NSC and UCD size-luminosity relation \citep[see Fig.\,13 in][]{georgiev14} while similarly comparing unfavourably with extremely extended Galactic globular clusters such as NGC\,2419.

Finally, while contamination of HCG90-DW5 by foreground sources makes the structural parameter estimation of its spheroid component challenging, constructing a good mask for the surrounding objects allowed us to find a stable solution that is consistent across both filters (see Fig.\,\ref{fig:dwarf_model}).  

\section{Discussion and Summary}\label{Sect:discussion}
We present the discovery of five new LSB dwarf galaxy candidates in HCG\,90 based on VLT/VIMOS $U-$ and $I-$band imaging.~Assuming distances concurrent with HCG\,90 ($D\!=\!33.1$\,Mpc), Table\,\ref{table:info} lists their photometric and morphological properties.~Three dwarfs show typical $r_{\rm eff}\!\approx\!1$\,kpc, while HCG90-DW3 and DW4 are smaller at $r_{\rm eff}\!\simeq\!700\!-\!850$\,pc.~These $r_{\rm eff}$ are consistent with those found by \cite{vdBurg16} for dwarfs in galaxy clusters in the sense that dwarfs at smaller cluster-centric radii are more compact with $r_{\rm eff}\la$1\,kpc due to harassment and tidal limitation by the hostile intra-cluster gravitational field. 

Fig.\,\ref{fig:lumsize} (left panel) shows $r_{\rm eff}$ as a function of the absolute $U$-band magnitude for the new candidates, compared to Fornax cluster dwarfs \citep[][Eigenthaler et al. 2016, {\it in prep.}]{munoz15} and the LG dwarf spheroidal (dSph) population, for which we estimate their $U$-band magnitudes based on the $V$-band luminosities and metallicities available in \cite{McCon12}.~With this information and assuming 12\,Gyr old stellar populations, we use \cite{bc03} models (hereafter BC03) to obtain the corresponding $U\!-\!V$ colours to convert the tabulated $V$-band luminosities in \cite{McCon12} to the corresponding $U$-band fluxes.~The new HCG\,90 dwarf galaxy candidates have luminosities in the range of $-11.65\!\le\!M_{U_{0}}\!\le\!-9.42$ and $-12.79\!\le\!M_{I_{0}}\!\le\!-10.58$\,mag, placing them in the magnitude regime of the average Fornax cluster dwarf, but with somewhat larger effective radii at a given luminosity.~This combination places the HCG\,90 dwarfs near the surface brightness limit of the Fornax dwarf sample.

The right panel of Fig.\,\ref{fig:lumsize} shows the $U_0$ vs.~$(U\!-\!I)_0$ colour-magnitude diagram (CMD) for the HCG90 and Fornax cluster dwarf galaxies.~Population synthesis model predictions from BC03 are shown for three different total stellar masses: 10$^6$, 10$^7$ and 10$^8\,M_{_\odot}$.~Despite the well-known age-metallicity-extinction degeneracy, with only two photometric bands our diagnostic capabilities are also degenerate with total stellar mass.~However, as the reddening vector primarily points parallel to the model isochrones, our main limitation lies in the age-metallicity-mass (AMM) degeneracy.~With this in mind, we estimate approximate ages and metallicities for the new dwarfs according to the model predictions. 

DW1, DW2, and DW4 are likely to be older than 1\,Gyr.~Their CMD parameters are consistent with metallicities $\lesssim\!0.02\,Z_\odot$ and ages $\gtrsim\!10$\,Gyr at $10^8\,M_\odot$ or super-solar metallicities for ages $\sim\!1$\,Gyr with SSP models scaled to $10^{6.5}\,M_\odot$.~DW3 and DW5 have bluer $(U\!-\!I)_0$ colours and given all the AMM degeneracies, their colours and luminosities are consistent with ages $\lesssim\!2$\,Gyr, unless their stellar populations are metal-free, in which case they may be a few Gyr older.~If DW3 is $\sim\!2$\,Gyr old its metallicity is $0.05\,Z_\odot$ with a total stellar mass of $10^{6.3}\,M_\odot$.~If its metallicity is solar or higher, its age estimate lies at $\sim\!500$\,Myr with $10^{5.7}\,M_\odot$. Similar arguments apply to the $2.2$\,mag brighter DW5, for which a mass of $10^{7.2}\,M_\odot$ is consistent with an age of $\sim\!2$\,Gyr and a metallicity of $0.05\,Z_\odot$.~If its metallicity is solar or above, it would have an age of $\sim\!500$\,Myr and a total stellar mass of $10^{6.3}\,M_\odot$.~However, we caution that due to the severe contamination of the DW5 surface brightness model fits (see Fig.\,\ref{fig:dwarf_model}), we consider its stellar population characterization to be uncertain.~In case of substantial internal reddening of the order $A_V=0.5$\,mag (see vector in the right panel of Fig.\,\ref{fig:lumsize}) ages would increase by a factor $\sim\!2\!-\!5$, leaving the metallicity and stellar mass estimates less affected.~Clearly, supplemental near-infrared imaging and/or spectroscopic observations are needed to provide more robust stellar population parameter constraints.

\begin{table}
\caption{Age, metallicity and mass (AMM) estimates for all of the HCG\,90 LSB dwarf candidates.~The two groups ($t_1,Z_1,M_{\star,1}$) and ($t_2,Z_2,M_{\star,2}$) show the stellar population parameter ranges, given the AMM degeneracy and measurement uncertainties (see text for details).~The last row gives the corresponding information for the central object in HCG90-DW4.\label{table:parameters}}
\centering
\setlength{\tabcolsep}{5pt} 
\renewcommand{\arraystretch}{1.3} 
\begin{tabular}{m{2cm} ccm{0.5cm} ccc}
\hline
\hline
ID      &  $t_1$      & $Z_1$               &$M_{\star,1}$  & $t_2$       & $Z _2$              & $M_{\star,2}$  \\
         &  [Gyr] & [Z$_\odot$] &[$M_\odot$] & [Gyr]   & [Z$_\odot$] & [$M_\odot$]  \\
\hline
\vspace{0.13cm}
HCG90-DW1 &    $\geq$\,10 & $\leq$\,0.02 & 10$^{8.0}$ & $\simeq$\,1 &  2.5 &  10$^{6.5}$ \\
HCG90-DW2 &   $\geq$\,10  & $\leq$\,0.02 & 10$^{8.0}$ & $\simeq$\,1 & 2.5 & 10$^{6.5}$\\
HCG90-DW3 &   $\simeq$\,2 & $\leq$\,0.05 & 10$^{6.3}$ & $\simeq$\,0.5 & > 1 & 10$^{5.7}$\\
HCG90-DW4 &   $\geq$\,10 & $\leq$\,0.02 & 10$^{8.0}$ & $\simeq$\,1 &  2.5 & 10$^{6.5}$\\
HCG90-DW5 &   $\simeq$\,2 & $\leq$0.05 &  10$^{7.2}$ & $\simeq$\,0.5 &  > 1 & 10$^{6.3}$ \\
\hline
HCG90-DW4n  &  $\geq$\,10 & $\leq$0.02 &  10$^{6.5}$ & 0.7\,-\,1 &  >1 &  10$^{4.9}$ \\

\end{tabular}
\end{table}

Assuming ages and metallicities as predicted by the models from the CMD, we estimate their stellar masses, $M_*$, from the BC03 mass-to-light ratios adopting two metallicities: $0.02\,Z_\odot$ and $Z_\odot$.~Our measured colours and luminosities for the spheroid components correspond to stellar mass ranges of $M_*|_{0.02Z_\odot}\!\simeq\!10^{6.3}\!-\!10^8\,M_{\odot}$ and $M_*|_{Z_\odot}\!\simeq\!10^{5.7}\!-\!10^{6.5}\,M_{\odot}$, with DW3 and DW2 being the least and most massive dwarfs, respectively.~For the nuclear object DW4n, we estimate $M_*|_{0.02Z_\odot}\!\simeq\!10^{6.5}$ and $M_*|_{Z_\odot}\!\simeq\!10^{4.9}\,M_{\odot}$.~A summary of the parameters can be found in Table~\ref{table:parameters}.

All new LSB dwarf candidates are located within (projected) 70\,kpc of the closest giant galaxy, although the limited VIMOS field-of-view makes it entirely possible that yet more dwarfs exist at larger radii.~These properties, combined with their morphologies, resemble dSphs found in the LG (see Fig.\,\ref{fig:lumsize}) with closest LG dSph analogs being Fornax with $r_{\rm eff}\!=\!710$\,pc (though with higher surface brightness, $\mu_{V,{\rm eff}}\!=\!24$\,mag\,arcsec$^{-2}$), Cetus ($r_{\rm eff}\!=\!710$\,pc and $\mu_{V,{\rm eff}}\!=\!26$\,mag\,arcsec$^{-2}$), Andromeda I, II and XXIII with \citep[$r_{\rm eff}\!=\!672$,\,1176,\,1029\,pc and $\mu_{V,{\rm eff}}\!=\! 25.8$,\,26.3,\,27.8\,mag\,arcsec$^{-2}$, respectively; see][]{McCon12}.~Notably, the LG dSphs have similar stellar masses of a few $10^6\!-\!10^7\,M_{\odot}$ compared to the HCG\,90 dwarf candidates in this work.

The case of HCG\,90-DW2 and DW3 is interesting in that the light profile modelling solution strongly favours two distinct components seen closely in projection (see centre-panel of Fig.\,\ref{fig:dwarf_model}).~While we cannot reject the possibility that DW2/3 is a single galaxy with an irregular shape, we may be observing a binary pair of LSB dwarfs with a projected separation of $<\!2$\,kpc.~Dwarf binaries such as DW2/3 may not be unprecedented in the local universe.~The most similar known systems are perhaps the LMC/SMC pair in the LG or the NGC\,4681/4625 association reported by \cite{pearson16}, both with much greater separations of 11 and (projected) 9.2\,kpc, respectively.~Intriguingly, DW2/3 are similar in both structural parameters and on-sky separation to those reported by \cite{crno14}, which together lie $\sim\!90$\,kpc from their own giant host NGC\,5128.~In any case, if this pair is truly physically associated, then it implies that dwarf-dwarf interactions may not be uncommon in dense galactic environments.

The nearest dwarf to the core of the compact group is HCG90-DW5 with potentially interesting properties.~It is the brightest candidate with the bluest $(U-I)_0$\,=\,1.14\,mag colour index.~Given its stellar population properties (see discussion above) and considering its close projected proximity to the three interacting giants, we suggest that this object may be a TDG arising from the tidal debris of the interacting giants.~Noting the above, this interpretation is hampered by the contamination of DW5's light profile and will require careful follow-up observations to confirm.

The spheroid component of HCG90-DW4 appears to host a central compact object (DW4n) with $r_{\rm eff}\!\approx\!46\!-\!63$\,pc.~However, its properties do not agree with those of a nuclear star cluster, which are generally about $5\!-\!10\times$ more compact and brighter than DW4n \citep[e.g.][]{georgiev14}.~Its size is marginally resolved by our data. This is illustrated by its radial profile in Fig.~\ref{fig:dwarf4}, which is slightly more extended than a stellar PSF.~An alternative explanation is that DW4n could represent a so-called nuclear stellar disc (NSD).~NSDs are typically located in the core regions of their host galaxies and have $r_{\rm eff}$ of 10s to 100s of pc \citep{vdbosch94} and are regularly found in early-type galaxies, but rarely in spirals \citep{piz02,ledo10}.~The shaded region of Fig.\ref{fig:lumsize} (left panel) shows the parameter space of a sample of nuclear discs studied by \cite{ledo10} where DW4n is slightly off from the faintest/smallest area of this range.~DW4n's colour and luminosity are consistent with old ages $>\!10$\,Gyr and metallicities $0.02\,Z_\odot$ at a stellar mass of $10^{6.5}\,M_\odot$ or younger ages $\sim\!0.7\!-\!1$\,Gyr and super-solar metallicities with a total stellar mass of $10^{4.9}\,M_\odot$ (see Table\,\ref{table:parameters}).  

On the other hand, if DW4n is a background source, then at 100\,Mpc distance (e.g.\ Coma cluster), it would have $r_{\rm eff}\!\simeq\!170$\,pc and $M_{U,0}\!\simeq\!-9.82$\,mag, and at 750\,Mpc (e.g.\ Abell\,1689 cluster) this would increase further to $r_{\rm eff}\!\simeq\!1.3$\,kpc and $M_{U,0}\!\simeq\!-14.22$\,mag.~In these cases, HCG\,90-DW4n falls on the faintest regions of the size-luminosity relation of early-type galaxies (see Fig.~\ref{fig:lumsize}).~Nonetheless, if it were an early-type galaxy we would expect a higher S\'ersic index (e.g.\ $n\gtrsim1$), consistent with massive ellipticals.~With the current data we cannot confirm the nature of this object or its distance, and despite a lack of observational evidence for NSDs in dwarf galaxies, if DW4 is a member of HCG\,90, then its structural parameters make it a good candidate for the first such object found inside a faint and relatively compact LSB host. Spectroscopic follow-up observations of DW4n should verify its HCG90 membership and, thereby, shed more light on its true nature.

\section*{Acknowledgements}

We thank the referee, Tyler Desjardins, for a constructive referee report that help to improve the clarity of the presentation of our results.~Y.O.-B.~acknowledges financial support through CONICYT-Chile (grant CONICYT-PCHA/Doctorado Nacional/2014-21140651). M.A.T.~acknowledges the financial support through an excellence grant from the "Vicerrector\'ia de Investigaci\'on", the Institute of Astrophysics Graduate School Fund at Pontificia Universidad Cat\'olica de Chile and the European Southern Observatory Graduate Student Fellowship program. T.H.P.~acknowledges the support through a FONDECYT Regular Project Grant (No.~1161817) and BASAL Center for Astrophysics and Associated Technologies (PFB-06).~This research has made use of the NASA Astrophysics Data System Bibliographic Services and the NASA Extragalactic Database.~Software packages used in the analysis include {\sc Python/NumPy} v.1.9.1 and {\sc Python/Scipy} v0.15.1 \citep[][]{jon01,van11}, {\sc Python/astropy} \citep[v1.0.1;][]{ast13}, and {\sc Python/matplotlib} \citep[v1.4.2;][]{hun07} as well as the {\sc astromatic} suite.












\bsp	
\label{lastpage}
\end{document}